# Deriving Z[J] from the time evolution operator

Dan Solomon
University of Illinois
Chicago, IL
dsolom2@uic.edu
May 30, 2014

## Abstract.

An important quantity in quantum field theory is the vacuum-to-vacuum transition amplitude in the presence of an external source. This quantity is often designated by $Z[J]$ and is the generator for the n-point Greens functions. In textbooks $Z[J]$ is often derived using a path integral formulation, however it will be shown here that it is possible to derive $Z[J]$ directly from the time evolution operator without using the path integral formulation.

## 1. Introduction.

An important quantity in quantum field theory is $Z[J]$ which is formally defined [1] as the vacuum-to-vacuum transition amplitude in the presence of an external source $J$,

$$Z[J]=\langle 0,\infty|0,-\infty\rangle_J \tag{1.1}$$

It can be shown that $Z[J]$ is the generator of the n-point Greens functions,

$$G(x_1,x_2,\ldots,x_n)=\frac{1}{i^n}\frac{\delta^n Z[J]}{\delta J(x_1)\delta J(x_2)\ldots\delta J(x_n)}\bigg|_{J=0} \tag{1.2}$$

which are instrumental in the calculation of scattering amplitudes.

In many textbooks $Z[J]$ is derived using the path integral approach (See, for example, [1,2,3,4,5]). In this paper I will demonstrate that $Z[J]$ can be derived directly from the time evolution operator without using the path integral approach. We will also derive an expression for the S-matrix operator which will be designated by $\hat{S}$. It will also be shown how the path integral formulation of the theory can be recovered by taking the inverse Fourier Transform of $Z[J]$.

Throughout the discussion we will assume $\hbar = c = 1$ . Also we will work in 3-1 dimensions where $x \to (\vec{x}, t)$ In the following discussion the model system that will be considered is a neutral spin-zero scalar field with $\varphi^4$ coupling . The classical action for this system is given by,

$$S = \int dt \int d^3x \left\{ \frac{1}{2}\varphi\left( -\frac{\partial^2}{\partial t^2} + \frac{\partial^2}{\partial x^2} - m^2 \right)\varphi - \frac{\lambda}{4!}\varphi^4 \right\} \tag{1.3}$$

where $m$ is the mass.

## 2. The interaction picture

In this section we will briefly review some properties of the interaction picture formulation of quantum field theory and derive the evolution operator $\hat{U}_I(t, t_0)$ for our model system. (For a more detailed discussion see [2] or [6]). In the interaction picture the time evolution of the system is divided between the field operator and the state vector. Let $|\Omega_I(t_0)\rangle$ be the state vector of the system at some initial time $t_0$ and $|\Omega_I(t_1)\rangle$ the state vector at some later time $t_1$. The relationship between the two states is,

$$|\Omega_I(t_1)\rangle = \hat{U}_I(t_1, t_0)|\Omega_I(t_0)\rangle \tag{2.1}$$

where $\hat{U}_I(t_1, t_0)$ is the unitary time evolution operator. The subscript $I$ in the above expression is used to indicate that we are working in the interaction picture.

In order to determine $\hat{U}_I(t_1, t_0)$ we will start in the Schrodinger picture. In this case the field operators $\hat{\varphi}_S(\vec{x})$ are time-independent and the time dependent state vector $|\Omega_S(t)\rangle$ obeys the Schrodinger equation,

$$i\frac{\partial}{\partial t}|\Omega_S\rangle = \hat{H}^S|\Omega_S\rangle \tag{2.2}$$

where $\hat{H}^S$ is the Hamiltonian operator in the Schrodinger picture.

For our model system $\hat{H}^S = \hat{H}_0^S + H_1^S$ where,

$$\hat{H}_0^S = \int \frac{d^3p}{(2\pi)^3}\hat{a}_p^\dagger \hat{a}_p \omega_p \quad \text{and} \quad H_1^S = \frac{\lambda}{4!}\int \hat{\varphi}_S^4(\vec{x}) d^3x \tag{2.3}$$

where $\omega_{\vec{p}} = \sqrt{|\vec{p}|^2 + m^2}$ and $\hat{a}_{\vec{p}}^{\dagger}$ and $\hat{a}_{\vec{p}}$ are the creation and destruction operators. They obey the usual commutation relationships,

$$\left[\hat{a}_{\vec{p}}, \hat{a}_{\vec{q}}^{\dagger}\right] = (2\pi)^3 \delta^3(\vec{p} - \vec{q}); \quad \left[\hat{a}_{\vec{p}}, \hat{a}_{\vec{q}}\right] = \left[\hat{a}_{\vec{p}}^{\dagger}, \hat{a}_{\vec{q}}^{\dagger}\right] = 0 \tag{2.4}$$

$\hat{\varphi}_S(\vec{x})$ is the time-independent Schrodinger field operator and is given by,

$$\hat{\varphi}_S(\vec{x}) = \int \frac{d^3 p}{(2\pi)^3} \sqrt{\frac{1}{2\omega_{\vec{p}}}} \left(\hat{a}_{\vec{p}}^{\dagger} e^{-i\vec{p}\cdot\vec{x}} + \hat{a}_{\vec{p}} e^{+i\vec{p}\cdot\vec{x}}\right) \tag{2.5}$$

Define the vacuum state $|0\rangle$ as the state which satisfies,

$$\langle 0|0\rangle = 1; \ \hat{a}_{\vec{p}}|0\rangle = 0 \text{ and } \langle 0|\hat{a}_{\vec{p}}^{\dagger} = 0 \text{ for all } \vec{p} \tag{2.6}$$

$|0\rangle$ is the vacuum state for the unperturbed system, that is, the system when the coupling $\lambda = 0$.

Switch to the interaction picture as follows. Define the interaction state vector $|\Omega_I\rangle$ by,

$$|\Omega_I\rangle = \exp\left(i\hat{H}_0^S t\right)|\Omega_S\rangle \tag{2.7}$$

This, of course, yields $|\Omega_S\rangle = \exp\left(-i\hat{H}_0^S t\right)|\Omega_I\rangle$ which can be substituted into (2.2) to obtain,

$$i\frac{\partial}{\partial t}|\Omega_I\rangle = \hat{H}_1^I|\Omega_I\rangle \tag{2.8}$$

where $\hat{H}_1^I = \exp\left(+i\hat{H}_0^S t\right)\hat{H}_1^S \exp\left(-i\hat{H}_0^S t\right)$. Use this along with (2.3), (2.4), and (2.5) to obtain,

$$\hat{H}_1^I(t) = \frac{\lambda}{4!}\int \hat{\varphi}_I^4(\vec{x}, \mathrm{t}) d^3 x \tag{2.9}$$

where the interaction field operator $\hat{\varphi}_I(\vec{x}, \mathrm{t})$ is given by,

$$\hat{\varphi}_I(\vec{x}, \mathrm{t}) = e^{i\hat{H}_0^S t}\hat{\varphi}_S(\vec{x}) e^{-i\hat{H}_0^S t} = \int \frac{d^3 p}{(2\pi)^3}\sqrt{\frac{1}{2\omega_{\vec{p}}}}\left(\hat{a}_{\vec{p}}^{\dagger} e^{+i\omega_p t} e^{-i\vec{p}\cdot\vec{x}} + \hat{a}_{\vec{p}} e^{-i\omega_p t} e^{+i\vec{p}\cdot\vec{x}}\right) \tag{2.10}$$

It is readily shown that $\hat{\varphi}_I(\vec{x}, \mathrm{t})$ satisfies,

$$\left(\frac{\partial^2}{\partial t^2}-\vec{\nabla}^2+m^2\right)\hat{\varphi}_I\left(\vec{x},\mathrm{t}\right)=0 \tag{2.11}$$

From (2.1) and (2.8) it can be seen that $\hat{U}_I\left(t,t_0\right)$ obeys the equation,

$$i\frac{\partial}{\partial t}\hat{U}_I\left(t,t_0\right)=\hat{H}_1^I\hat{U}_I\left(t,t_0\right) \tag{2.12}$$

with the initial condition $\hat{U}_I\left(t_0,t_0\right)=1$. It can readily be shown that [2],

$$\hat{U}_I\left(t_1,t_0\right)=T\exp\left(-i\int_{t_0}^{t_1}H_1^I\left(t\right)dt\right) \tag{2.13}$$

where $T$ stands for the time-ordering operator. It has the effect of moving the field operators $\hat{\varphi}_I\left(\vec{x},\mathrm{t}\right)$ that occur at earlier times to the right. For example,

$$T\left(\hat{\varphi}_I\left(\vec{x},\mathrm{t}\right)\hat{\varphi}_I\left(\vec{x}',\mathrm{t}'\right)\right)=\left\langle\begin{matrix}\hat{\varphi}_I\left(\vec{x},\mathrm{t}\right)\hat{\varphi}_I\left(\vec{x}',\mathrm{t}'\right)\text{ if } t>t'\\ \hat{\varphi}_I\left(\vec{x}',\mathrm{t}'\right)\hat{\varphi}_I\left(\vec{x},\mathrm{t}\right)\text{ if } t'>t\end{matrix}\right. \tag{2.14}$$

Eq. (2.13) is actually a symbolic representation of the following expression,

$$\hat{U}_I\left(t_1,t_0\right)=1+\sum_{n=1}^{\infty}\frac{\left(-i\right)^n}{n!}\int_{t_0}^{t_1}dt_1'\int_{t_0}^{t_1}dt_2'\ldots\int_{t_0}^{t_1}dt_n'T\left(H_1^I\left(t_1'\right)H_1^I\left(t_2'\right)\ldots H_1^I\left(t_n'\right)\right) \tag{2.15}$$

Use (2.9) in (2.13) to obtain,

$$\hat{U}_I\left(t_1,t_0\right)=T\exp\left(-i\frac{\lambda}{4!}\int_{t_0}^{t_1}\hat{\varphi}_I^4dx\right) \tag{2.16}$$

where to simplify the expressions we have replaced $\int_{t_0}^{t_1}dt\int f\left(\vec{x},t\right)d^3x$ with $\int_{t_0}^{t_1}f\left(x\right)dx$. This convention will be used in the rest of the discussion.

## 3. Derivation of Z[J].

In this section it will be shown that we can derive $Z\left[J\right]$ from the time evolution operator ,$\hat{U}_I\left(t_1,t_0\right)$. It can be shown that,

$$\frac{1}{i}\frac{\delta}{\delta J(x_1)}\frac{1}{i}\frac{\delta}{\delta J(x_2)}\cdots\frac{1}{i}\frac{\delta}{\delta J(x_n)}\mathrm{T}\exp\left(i\int_{-\infty}^{+\infty}dxJ\hat{\varphi}_I\right)=T\left\{\hat{\varphi}_I(x_1)\hat{\varphi}_I(x_2)\ldots\hat{\varphi}_I(x_n)\exp\left(i\int_{-\infty}^{+\infty}dxJ\hat{\varphi}_I\right)\right\} \quad (3.1)$$

where $J(x)$ is some real-valued function. Define the expression,

$$\hat{U}_M(t_1,t_0;J)=\exp\left(-i\int_{t_0}^{t_1}dx\frac{\lambda}{4!}\left(\frac{1}{i}\frac{\delta}{\delta J(x)}\right)^4\right)T\exp\left(i\int_{-\infty}^{+\infty}J(x)\hat{\varphi}_I(x)dx\right) \quad (3.2)$$

Using (3.1) in the above we obtain,

$$\hat{U}_M(t_1,t_0;J)=T\left(\exp\left(-i\frac{\lambda}{4!}\int_{t_0}^{t_1}\hat{\varphi}_I^4dx\right)\exp\left(i\int_{-\infty}^{+\infty}J(x)\hat{\varphi}_I(x)dx\right)\right) \quad (3.3)$$

Now let us try to give some physical meaning to $\hat{U}_M(t_1,t_0;J)$. Comparing (3.3) to (2.16) we obtain,

$$\hat{U}_I(t_1,t_0)=\hat{U}_M(t_1,t_0;J)\Big|_{J=0} \quad (3.4)$$

Therefore $\hat{U}_M(t_1,t_0;J)\Big|_{J=0}$ is an alternate expression for the time evolution operator $\hat{U}_I(t_1,t_0)$. Now what, if any, is the physical meaning of the function $J(x)$. Refer to (3.3) and assume that $J(x)$ is only non-zero for time $t$ such that $t_1\ge t\ge t_0$ then we can write,

$$\hat{U}_M(t_1,t_0;J)=T\left(\exp\left(-i\frac{\lambda}{4!}\int_{t_0}^{t_1}\hat{\varphi}_I^4dx\right)\exp\left(i\int_{t_0}^{t_1}J(x)\hat{\varphi}_I(x)dx\right)\right) \quad (3.5)$$

It is shown in Appendix 1 that this can be rewritten as,

$$\hat{U}_M(t_1,t_0;J)=T\exp\left(-i\frac{\lambda}{4!}\int_{t_0}^{t_1}\hat{\varphi}_I^4dx+i\int_{t_0}^{t_1}J(x)\hat{\varphi}_I(x)dx\right)=T\exp\left(-i\int_{t_0}^{t_1}\left(\frac{\lambda}{4!}\hat{\varphi}_I^4-J(x)\hat{\varphi}_I(x)\right)dx\right) \quad (3.6)$$

This expression is the time evolution operator for our model system in the presence of an external source $J(x)$ where the effect of the source is to add the term $\int dxJ(x)\varphi(x)$ to the original action which was specified by (1.3). Therefore, under the conditions that $J(x)$ is only non-zero for time $t$ such that $t_1\ge t\ge t_0$, $\hat{U}_M(t_1,t_0;J)$ can be interpreted as the Interaction Picture evolution operator in the presence of an external source. However we have defined $\hat{U}_M(t_1,t_0;J)$ according to (3.3) and the physical meaning attributed to $J$ for times outside of the range $t_1\ge t\ge t_0$ is obscure, therefore in general we will assume $J(x)$ to be a mathematical device and not concern ourselves too much with its physical meaning.

Using Wicks theorem It is shown in Appendix 2 that,

$$T\exp\left(i\int_{-\infty}^{+\infty}J(x)\hat{\varphi}_I(x)dx\right)=:\exp\left(i\int_{-\infty}^{+\infty}J(x)\hat{\varphi}_I(x)dx\right):\exp\left(-\frac{i}{2}\int_{-\infty}^{+\infty}\int_{-\infty}^{+\infty}J(x)\Delta_F(x-x')J(x')dxdx'\right) \quad (3.7)$$

where $\Delta_F(x-x')=-iT\langle 0|\hat{\varphi}_I(x)\hat{\varphi}_I(x')|0\rangle$ is the Feynman propagator . The colons : : in the above expression stand for the normal order operation. This means that any expression between the colons containing combinations of creation and destruction operators the destruction operator are all moved to the right, e.g., $:a_k^\dagger\hat{a}_q a_p^\dagger:=a_k^\dagger a_p^\dagger\hat{a}_q$ .

From Section 6.1 of [1] the Feynman propagator is given by,

$$\Delta_F(x-x')=\frac{1}{(2\pi)^4}\int d^4k\left(\frac{e^{-ik(x-x')}}{k^2-m^2+i\varepsilon}\right) \quad (3.8)$$

where $\varepsilon\to 0$ from above. $\Delta_F(x-x')$ satisfies,

$$K_x\Delta_F(x-x')=-\delta^4(x-x') \quad (3.9)$$

where,

$$K_x=\left(\frac{\partial^2}{\partial t^2}-\vec{\nabla}^2+m^2-i\varepsilon\right) \quad (3.10)$$

Use (3.7) in (3.2) to obtain,

$$\hat{U}_M(t_1,t_0;J)=\exp\left(-i\int_{t_0}^{t_1}dx\frac{\lambda}{4!}\left(\frac{1}{i}\frac{\delta}{\delta J(x)}\right)^4\right)\left[\begin{array}{l}:\exp\left(i\int_{-\infty}^{+\infty}J(x)\hat{\varphi}_I(x)dx\right):\\ \times\exp\left(-\frac{i}{2}\int_{-\infty}^{+\infty}\int_{-\infty}^{+\infty}J(x)\Delta_F(x-x')J(x')dxdx'\right)\end{array}\right] \quad (3.11)$$

Now consider the quantity $\langle 0|\hat{U}_M(t_1,t_0;J)|0\rangle$ . Use the fact that $\langle 0|:\hat{\varphi}_I(x_1)\hat{\varphi}_I(x_2)\ldots\hat{\varphi}_I(x_n):|0\rangle=0$ to obtain,

$$\langle 0|:\exp\left(i\int_{t_0}^{t_1}J(x)\hat{\varphi}_I(x)dx\right):|0\rangle=1 \quad (3.12)$$

Use this in (3.11) to obtain,

$$\langle 0|\hat{U}_M(t_1,t_0;J)|0\rangle = \exp\left(-i\int_{t_0}^{t_1} dx \frac{\lambda}{4!}\left(\frac{1}{i}\frac{\delta}{\delta J(x)}\right)^4\right)\exp\left(-\frac{i}{2}\int_{-\infty}^{+\infty}\int_{-\infty}^{+\infty} J(x)\Delta_F(x-x')J(x')dxdx'\right) \quad (3.13)$$

Now let $t_1 \to \infty$ and $t_0 \to -\infty$ to obtain,

$$\langle 0|\hat{U}_M(\infty,-\infty;J)|0\rangle = \exp\left(-i\int_{-\infty}^{\infty} dx \frac{\lambda}{4!}\left(\frac{1}{i}\frac{\delta}{\delta J(x)}\right)^4\right)\exp\left(-\frac{i}{2}\int_{-\infty}^{\infty}\int_{-\infty}^{\infty} J(x)\Delta_F(x-x')J(x')dxdx'\right) \quad (3.14)$$

This is identical to the non-normalized $Z[J]$ for $\varphi^4$ theory in the presence of an external source (See Eq. 6.82 of [1]). Therefore we can write,

$$Z[J] = N_r \langle 0|\hat{U}_M(\infty,-\infty;J)|0\rangle \quad (3.15)$$

where $N_r$ is given by,

$$N_r = \frac{1}{\langle 0|\hat{U}_M(\infty,-\infty;J)|0\rangle\Big|_{J=0}} = \frac{1}{\langle 0|\hat{U}_I(\infty,-\infty)|0\rangle} \quad (3.16)$$

where we have used (3.4) to obtain the second equality. The factor $N_r$ is necessary so that that $Z[J]\big|_{J=0} = 1$

## 4. S-matrix.

It will be shown that the $\hat{S}$ operator can also be obtained from $\hat{U}_M(t_1,t_0;J)$. From Section 6.8 of [1] the S-matrix is given by,

$$\hat{S} = :\exp\left(\int_{-\infty}^{+\infty}\hat{\varphi}_I(x)K_x\frac{\delta}{\delta J(x)}dx\right): Z[J]\Big|_{J=0} \quad (4.1)$$

Using (3.15) this can be written as, (4.2)

$$\hat{S} = N_r :\exp\left(\int_{-\infty}^{+\infty}\hat{\varphi}_I(x)K_x\frac{\delta}{\delta J(x)}dx\right):\langle 0|\hat{U}_M(\infty,-\infty;J)|0\rangle\Big|_{J=0} \quad (4.3)$$

Using (3.14) in the above and rearranging terms we get,

$$\hat{S}[J] = N_r \exp\left(-i\int_{-\infty}^{+\infty} dx \frac{\lambda}{4!}\left(\frac{1}{i}\frac{\delta}{\delta J(x)}\right)^4\right)\hat{Q}[J] \tag{4.4}$$

where for the moment we have dropped the requirement that $J = 0$ and where,

$$\hat{Q}[J] =: \exp\left(i\int_{-\infty}^{+\infty} \hat{\varphi}_I(x) K_x \frac{\delta}{i\delta J(x)} dx\right) \exp\left(-\frac{i}{2}\int_{-\infty}^{\infty}\int_{-\infty}^{\infty} J(x)\Delta_F(x-x')J(x')dxdx'\right): \tag{4.5}$$

Define $\hat{C} = \left(\int_{-\infty}^{+\infty} \hat{\varphi}_I(x) K_x \frac{\delta}{i\delta J(x)} dx\right)$ and $D = \int_{-\infty}^{\infty}\int_{-\infty}^{\infty} J(x)\Delta_F(x-x')J(x')dxdx'$ . Therefore,

$$\hat{Q}[J] =: \exp\left(i\hat{C}\right)\exp\left(-\frac{i}{2}D\right): \tag{4.6}$$

Expand the first exponential to obtain,

$$\hat{Q}[J] =: \left(1 + i\hat{C} + \frac{\left(i\hat{C}\right)^2}{2!} + \frac{\left(i\hat{C}\right)^3}{3!} + \ldots\right)\exp\left(-\frac{i}{2}D\right): \tag{4.7}$$

We will evaluate this term by term as follows. First, note that,

$$K_x \frac{\delta D}{\delta J(x)} = 2\int_{-\infty}^{\infty} K_x \Delta_F(x-x')J(x')dx' = -2J(x) \tag{4.8}$$

where we have used $K_x \Delta_F(x-x') = -\delta^4(x-x')$ . Use this to obtain,

$$K_x \frac{\delta}{i\delta J(x)} \exp\left(-\frac{i}{2}D\right) = J(x)\exp\left(-\frac{i}{2}D\right) \tag{4.9}$$

Therefore,

$$\hat{C}\exp\left(-\frac{i}{2}D\right) = \hat{B}\exp\left(-\frac{i}{2}D\right) \tag{4.10}$$

where ,

$$\hat{B} = \int_{-\infty}^{+\infty} J\hat{\varphi}_I dx \tag{4.11}$$

Next evaluate the next term in (4.7) which is $\left(i\hat{C}\right)^2 \exp\left(-\frac{i}{2}D\right)\Big/2!$ . From the above we obtain,

$$\hat{C}^2 \exp\left(-\frac{i}{2}D\right) = \hat{C}\hat{B}\exp\left(-\frac{i}{2}D\right) = \left(i\int_{-\infty}^{+\infty} \hat{\varphi}_I\left(x\right)K_x \frac{\delta}{i\delta J\left(x\right)}dx\right)\hat{B}\exp\left(-\frac{i}{2}D\right) \qquad (4.12)$$

To evaluate this use,

$$K_x \frac{\delta}{i\delta J\left(x\right)}\hat{B}\exp\left(-\frac{i}{2}D\right) = \left(K_x \frac{\delta}{i\delta J\left(x\right)}\hat{B}\right)\exp\left(-\frac{i}{2}D\right) + \hat{B}K_x \frac{\delta}{i\delta J\left(x\right)}\exp\left(-\frac{i}{2}D\right) \qquad (4.13)$$

Next, use (4.9) in the above to obtain,

$$K_x \frac{\delta}{i\delta J\left(x\right)}\hat{B}\exp\left(-\frac{i}{2}D\right) = \left(K_x \frac{\delta}{i\delta J\left(x\right)}\hat{B}\right)\exp\left(-\frac{i}{2}D\right) + \hat{B}J\left(x\right)\exp\left(-\frac{i}{2}D\right) \qquad (4.14)$$

Use (2.11) and (3.10) to obtain,

$$\left(K_x \frac{\delta}{i\delta J\left(x\right)}\hat{B}\right) = K_x \frac{\delta}{i\delta J\left(x\right)}\int_{-\infty}^{+\infty} J\hat{\varphi}_I dx = K_x\hat{\varphi}_I\left(x\right) = -i\varepsilon\hat{\varphi}_I\left(x\right) \to 0 \qquad (4.15)$$

The last step is due to the fact that $\varepsilon \to 0$ .

As a result of the above steps we obtain,

$$\hat{C}^2 \exp\left(-\frac{i}{2}D\right) = \int_{-\infty}^{+\infty} dx\hat{\varphi}_I\left(x\right)\hat{B}J\left(x\right)\exp\left(-\frac{i}{2}D\right) = \left(\int_{-\infty}^{+\infty} J\hat{\varphi}_I dx\right)\hat{B}\exp\left(-\frac{i}{2}D\right) \qquad (4.16)$$

Next use (4.11) to obtain,

$$\hat{C}^2 \exp\left(-\frac{i}{2}D\right) = \hat{B}^2 \exp\left(-\frac{i}{2}D\right) \qquad (4.17)$$

Continuing on it can be shown that,

$$\hat{C}^n \exp\left(-\frac{i}{2}D\right) = \hat{B}^n \exp\left(-\frac{i}{2}D\right) \qquad (4.18)$$

Therefore,

$$\hat{Q}\left[J\right] =: \exp\left(i\hat{C}\right)\exp\left(-\frac{i}{2}D\right) :=: \exp\left(i\hat{B}\right)\exp\left(-\frac{i}{2}D\right): \qquad (4.19)$$

The result is that,

$$\hat{S}[J] = N_r \exp\left(-i\int_{-\infty}^{+\infty} dx \frac{\lambda}{4!}\left(\frac{1}{i}\frac{\delta}{\delta J(x)}\right)^4\right)\left\{\begin{array}{l} :\exp\left(\int_{-\infty}^{+\infty} J\hat{\varphi}_I dx\right): \\ \times\exp\left(-\frac{i}{2}\int_{-\infty}^{\infty}\int_{-\infty}^{\infty} J(x)\Delta_F(x-x')J(x')dxdx'\right)\end{array}\right\} \quad (4.20)$$

Compare this result to (3.11) to obtain,

$$\hat{S}[J] = N_r \hat{U}_M(\infty, -\infty; J) \quad (4.21)$$

Therefore,

$$\hat{S} = N_r \hat{U}_M(\infty, -\infty; J)\Big|_{J=0} \quad (4.22)$$

## 5. Recovering the path integral approach.

As was stated in the Introduction most derivations of $Z[J]$ use the path integral approach. When this approach is used the expression for $Z[J]$ is,

$$Z[J] = Z[0]\int D[\varphi]\exp\left(iS[\varphi] + i\int_{-\infty}^{+\infty} J\varphi dx - \frac{\varepsilon}{2}\int_{-\infty}^{+\infty}\varphi^2 dx\right) \quad (5.1)$$

where $S[\varphi]$ is the classical action which, for our model system, is given by Eq. (1.3). (Note – do not confuse $\varphi$ with the field operators $\hat{\varphi}_I$ or $\hat{\varphi}_S$. In (5.1) $\varphi$ is simply a "dummy" variable that we integrate over).

In this paper we have found $Z[J]$ from the time evolution operator and and bypassed (5.1). It is the purpose of this section to show how the approach presented in this paper can also lead to the formula in (5.1). We do this by noticing that (5.1) is in the form of a Fourier transform. If we define,

$$\tilde{Z}[\varphi] = Z[0]\exp\left(iS[\varphi] - \frac{\varepsilon}{2}\int_{-\infty}^{+\infty}\varphi^2 dx\right) \quad (5.2)$$

then (5.1) can be written as,

$$Z[J] = \int D[\varphi]\tilde{Z}[\varphi]\exp\left(i\int_{-\infty}^{+\infty} J\varphi dx\right) \quad (5.3)$$

When written in this form it is apparent that $Z[J]$ is the Fourier transform of $\tilde{Z}[\varphi]$. Therefore if we already have determined $Z[J]$ we should be able to obtain $\tilde{Z}[\varphi]$ by taking the inverse Fourier transform of $Z[J]$. It will be shown that this is, indeed, the case. In the following we will take the inverse Fourier Transform of (3.15) and obtain (5.2). The result of this is that the path integral formulation can be recovered from $Z[J]$. Taking the inverse Fourier Transform of (3.15) we obtain,

$$\tilde{Z}[\varphi] = N_r \int D[J] \langle 0 | \hat{U}_M(\infty, -\infty; J) | 0 \rangle \exp\left( -i \int_{-\infty}^{+\infty} J\varphi dx \right) \tag{5.4}$$

What we want to determine is whether or not $\tilde{Z}[\varphi]$ as given by this equation will equal $\tilde{Z}[\varphi]$ as given by Eq. (5.2). It will be shown that this is the case.

Use (3.14) in the above to obtain,

$$\tilde{Z}[\varphi] = N_r \int D[J] \exp\left( -i \int_{-\infty}^{+\infty} J\varphi dx \right) \left\{ \begin{array}{l} \exp\left( -i \dfrac{\lambda}{4!} \displaystyle\int_{-\infty}^{+\infty} \left( \dfrac{i}{i} \dfrac{\delta}{\delta J(x)} \right)^4 dx \right) \\ \times \exp\left( -\dfrac{i}{2} \displaystyle\int_{-\infty}^{+\infty} J(x) \Delta(x - x') J(x') dx dx' \right) \end{array} \right\} \tag{5.5}$$

Integrate by parts to obtain,

$$\tilde{Z}[\varphi] = N_0 \int D[J] \exp\left( -i \int_{-\infty}^{+\infty} J\varphi dx \right) \left\{ \begin{array}{l} \exp\left( -i \dfrac{\lambda}{4!} \displaystyle\int_{-\infty}^{+\infty} \varphi^4 dx \right) \\ \times \exp\left( -\dfrac{i}{2} \displaystyle\int_{-\infty}^{+\infty} J(x) \Delta(x - x') J(x') dx dx' \right) \end{array} \right\} \tag{5.6}$$

In this step we have used $\frac{1}{i} \frac{\delta}{\delta J(x)} \exp\left( -i \int_{-\infty}^{+\infty} J\varphi dx \right) = -\varphi(x) \exp\left( -i \int_{-\infty}^{+\infty} J\varphi dx \right)$.

Rearrange terms to obtain,

$$\tilde{Z}[\varphi] = N_r \, \Upsilon[\varphi] \int D[J] \exp\left( -i \int_{-\infty}^{+\infty} J\varphi dx \right) \exp\left( -\frac{i}{2} \int_{-\infty}^{+\infty} J(x) \Delta(x - x') J(x') dx dx' \right) \tag{5.7}$$

where,

$$\Upsilon[\varphi] = \exp\left( -i \frac{\lambda}{4!} \int_{-\infty}^{+\infty} \varphi^4 dx \right) \tag{5.8}$$

Note that,

$$\frac{i}{2}\int_{-\infty}^{+\infty} J(x)\Delta(x-x')J(x')dxdx' + i\int_{-\infty}^{+\infty} J\varphi dx = \left(\begin{array}{l} \frac{i}{2}\int_{-\infty}^{+\infty}\left(J(x)-K_x\varphi(x)\right)\Delta(x-x')\left(J(x')-K_{x'}\varphi(x')\right)dxdx' \\ -\frac{i}{2}\int_{-\infty}^{+\infty} K_x\varphi(x)\Delta(x-x')K_{x'}\varphi(x')dxdx' \end{array}\right) \tag{5.9}$$

where we have used $K_x\Delta(x-x') = -\delta^4(x-x')$. Using this, the last part of the above equation can be written as,

$$\frac{i}{2}\int_{-\infty}^{+\infty} K_x\varphi(x)\Delta(x-x')K_{x'}\varphi(x')dxdx' = -\frac{i}{2}\int_{-\infty}^{+\infty}\varphi(x)K_x\varphi(x)dx \tag{5.10}$$

Use all this in (5.7) to obtain,

$$\tilde{Z}[\varphi] = N_r\,\Upsilon[\varphi]\exp\left(-\frac{i}{2}\int_{-\infty}^{+\infty}\varphi(x)K_x\varphi(x)dx\right)R[\varphi] \tag{5.11}$$

where,

$$R[\varphi] \equiv \int D[J]\exp\left(-\frac{i}{2}\int_{-\infty}^{+\infty}\left(J(x)-K_x\varphi(x)\right)\Delta(x-x')\left(J(x')-K_{x'}\varphi(x')\right)dxdx'\right) \tag{5.12}$$

Next, we will argue that $R[\varphi]$ is not dependent on $\varphi$ due to the fact that the change of the integration variable $J(x) \to J(x) - K_x\varphi(x)$ eliminates $\varphi$ from the integrand. Therefore $R[\varphi] \to R$ is just a constant.

The result of all this is that,

$$\tilde{Z}[\varphi] = \tilde{Z}[0]\exp\left(-i\int_{-\infty}^{+\infty}\left(\frac{1}{2}\varphi K_x\varphi + \frac{\lambda}{4!}\varphi^4\right)dx\right) \tag{5.13}$$

where all the "normalization" constants are absorbed in the term $\tilde{Z}[0]$. Use (3.10) in the above to obtain,

$$\tilde{Z}[\varphi] = \tilde{Z}[0]\exp\left(-i\int_{-\infty}^{+\infty}\left(\frac{1}{2}\varphi\left(\frac{\partial^2}{\partial t^2} - \vec{\nabla}^2 + m^2 - i\varepsilon\right)\varphi + \frac{\lambda}{4!}\varphi^4\right)dx\right) \tag{5.14}$$

Refer to (1.3) to obtain,

$$\tilde{Z}[\varphi]=\tilde{Z}[0]\exp\left(iS[\varphi]-\frac{\varepsilon}{2}\int_{-\infty}^{+\infty}\varphi^2 dx\right) \tag{5.15}$$

Compare this result to (5.2). Therefore the mathematical formulation associated with the path integral approach is recovered by taking the inverse Fourier Transform of $Z[J]$.

## 6. Conclusion.

In conclusion, we have considered a model scalar field as described in the Introduction. For this field we have obtained the interaction picture time evolution operator $\hat{U}_I(t_1,t_0)$. Next we define an operator $\hat{U}_M(t_1,t_0;J)$ by equation (3.2) and show that it is closely related to $\hat{U}_I(t_1,t_0)$. As was discussed $\hat{U}_I(t_1,t_0)=\hat{U}_M(t_1,t_0;J)\big|_{J=0}$ or, alternatively, in the situation where $J(\vec{x},t)$ is only non-zero when $t_0<t<t_1$ then $\hat{U}_I(t_1,t_0)=\hat{U}_M(t_1,t_0;J)$ if $J(\vec{x},t)$ is a source term and couples to the scalar field as discussed in Section 3. Having obtained $\hat{U}_M(t_1,t_0;J)$ we then obtain the published result for $Z[J]$ from Ref. [1] and show that $Z[J]=N_r\langle 0|\hat{U}_M(\infty,-\infty;J)|0\rangle$. A key step in this process was showing that (3.7) holds. The proof of this was given in Appendix 2. Next we examined the S-matrix operator as given in Ref. [1]. It was shown that $\hat{S}$ is related to $\hat{U}_M(\infty,-\infty;J)$ by Eq. (4.22). We have been able to derive these results without using the path integral formulation. Finally we show that we can perform a inverse Fourier Transform on $Z[J]$ to recover Eq. (5.1) which is the standard form that $Z[J]$ takes when the in path integral formulation of the problem is used.

## Appendix 1.

Show that (3.6) is true which is rewritten below for convenience,

$$T\exp\left(-i\frac{\lambda}{4!}\int_{t_0}^{t_1}\hat{\varphi}_I^4 dx+i\int_{t_0}^{t_1}J(x)\hat{\varphi}_I(x)dx\right)=T\exp\left(-i\int_{t_0}^{t_1}\left(\frac{\lambda}{4!}\hat{\varphi}_I^4-J(x)\hat{\varphi}_I(x)\right)dx\right) \tag{6.1}$$

Define,

$$\hat{A}(x)=-i\frac{\lambda}{4!}\hat{\varphi}_I^4 \text{ and } \hat{B}(x)=iJ(x)\hat{\varphi}_I(x) \tag{6.2}$$

and,

$$V_A = \int_{t_0}^{t} \hat{A}(x)\,dx \text{ and } V_B = \int_{t_0}^{t} \hat{B}(x)\,dx \tag{6.3}$$

Expand $T\exp(V_A + V_B)$ to obtain,

$$T\exp\left(\hat{V}_A + \hat{V}_B\right) = T\left(1 + \left(\hat{V}_A + \hat{V}_B\right) + \frac{1}{2!}\left(\hat{V}_A + \hat{V}_B\right)^2 + \frac{1}{3!}\left(\hat{V}_A + \hat{V}_B\right)^3 + h.o.t.\right) \tag{6.4}$$

where $h.o.t.$ stand for "higher order terms". Even though $\hat{V}_A$ and $\hat{V}_B$ may not commute the order does not matter because of the Time Ordering operator. For example,

$$T\left(\hat{V}_A\hat{V}_B\right) = T\left(\int_{t_0}^{t}\int_{t_0}^{t} \hat{A}(x)\,B(x')\,dxdx'\right) = T\left(\int_{t_0}^{t}\int_{t_0}^{t} B(x')\,\hat{A}(x)\,dxdx'\right) = T\left(\hat{V}_B\hat{V}_A\right) \tag{6.5}$$

Similarly we can show that $T\left(\hat{V}_A\hat{V}_B\hat{V}_B\right) = T\left(\hat{V}_B\hat{V}_A\hat{V}_B\right) = T\left(\hat{V}_B\hat{V}_B\hat{V}_A\right)$ etc. Using these results (6.4) becomes,

$$T\exp\left(\hat{V}_A + \hat{V}_B\right) = T\left(1 + \left(\hat{V}_A + \hat{V}_B\right) + \frac{1}{2!}\left(\hat{V}_A + 2\hat{V}_A\hat{V}_B + \hat{V}_B\right) + \frac{1}{3!}\left(\hat{V}_A^3 + 3\hat{V}_A^2\hat{V}_B + 3\hat{V}_A\hat{V}_B^2 + \hat{V}_B^3\right) + h.o.t.\right) \tag{6.6}$$

This can be written as,

$$T\exp\left(\hat{V}_A + \hat{V}_B\right) = T\left(1 + \left(\hat{V}_A + \hat{V}_B\right) + \frac{1}{2!}\left(\hat{V}_A + 2\hat{V}_A\hat{V}_B + \hat{V}_B\right) + \frac{1}{3!}\left(\hat{V}_A^3 + 3\hat{V}_A^2\hat{V}_B + 3\hat{V}_A\hat{V}_B^2 + \hat{V}_B^3\right) + h.o.t.\right) \tag{6.7}$$

Now compare this to $T\left(\exp\left(\hat{V}_A\right)\exp\left(\hat{V}_B\right)\right)$ . When this is expanded we obtain,

$$T\left(\exp\left(\hat{V}_A\right)\exp\left(\hat{V}_B\right)\right) = T\left(\left(1 + \hat{V}_A + \frac{1}{2!}\hat{V}_A^2 + \frac{1}{3!}\hat{V}_A^3 + h.ot.\right)\left(1 + \hat{V}_B + \frac{1}{2!}\hat{V}_B^2 + \frac{1}{3!}\hat{V}_B^3 + h.o.t\right)\right) \tag{6.8}$$

It can easily be seen that when taken to the corresponding order,

$$T\exp\left(\hat{V}_A + \hat{V}_B\right) = T\left(\exp\left(\hat{V}_A\right)\exp\left(\hat{V}_B\right)\right)$$

This has been demonstrated up to the 3rd order in the expansion but it should be evident that it will work to any order. Therefore (3.6) is correct.

## Appendix 2.

We want to show that (3.7) is true which, for convenience, is rewritten below,

$$T\exp\left(i\int_{-\infty}^{+\infty} J(x)\hat{\varphi}_I(x)\,dx\right) =: \exp\left(i\int_{-\infty}^{+\infty} J(x)\hat{\varphi}_I(x)\,dx\right) : \exp\left(-\frac{i}{2}\int_{-\infty}^{+\infty}\int_{-\infty}^{+\infty} J(x)\Delta_F(x-x')J(x')\,dx\,dx'\right)$$

(7.1)

In proving this we will use Wicks theorem as discussed in Section 8.5 of [2].

To simplify notation let $\int_{-\infty}^{+\infty} J(x)\hat{\varphi}_I(x)\,dx \to J^n\hat{\varphi}_n$. Using this convention expand the left side of (7.1) to obtain,

$$T\exp\left(iJ^n\hat{\varphi}_n\right) = T\sum_{m=0}^{\infty}\frac{i^m}{m!}\left(J^n\hat{\varphi}_n\right)^m = \sum_{m=0}^{\infty}\frac{i^m}{m!}J^{n_1}J^{n_2}\ldots J^{n_m}T\left(\hat{\varphi}_{n_1}\hat{\varphi}_{n_2}\ldots\hat{\varphi}_{n_m}\right) \tag{7.2}$$

Next we have to evaluate the time ordered products term by term using Wicks theorem . Per the discussion in [2],

$$T\left(\hat{\varphi}_{n_1}\hat{\varphi}_{n_2}\right) =: \hat{\varphi}_{n_1}\hat{\varphi}_{n_2} : + S_{n_1n_2} \tag{7.3}$$

where $S_{n_1n_2} = \langle 0|T\left(\hat{\varphi}_{n_1}\hat{\varphi}_{n_2}\right)|0\rangle$ . Next evaluate $T\left(\hat{\varphi}_{n_1}\hat{\varphi}_{n_2}\hat{\varphi}_{n_3}\right)$ to obtain,

$$T\left(\hat{\varphi}_{n_1}\hat{\varphi}_{n_2}\hat{\varphi}_{n_3}\right) =: \hat{\varphi}_{n_1}\hat{\varphi}_{n_2}\hat{\varphi}_{n_3} + S_{n_1n_2}\hat{\varphi}_{n_3} + S_{n_1n_3}\hat{\varphi}_{n_2} + S_{n_2n_3}\hat{\varphi}_{n_2} : \tag{7.4}$$

Since the indices, $n_i$ , are dummy indices then when $T\left(\hat{\varphi}_{n_1}\hat{\varphi}_{n_2}\hat{\varphi}_{n_3}\right)$ is used in (7.2) we can write,

$$T\left(\hat{\varphi}_{n_1}\hat{\varphi}_{n_2}\hat{\varphi}_{n_3}\right) \to : \hat{\varphi}_{n_1}\hat{\varphi}_{n_2}\hat{\varphi}_{n_3} : + 3\hat{\varphi}_{n_1}S_{n_2n_3} \tag{7.5}$$

Similarly for $T\left(\hat{\varphi}_{n_1}\hat{\varphi}_{n_2}\hat{\varphi}_{n_3}\hat{\varphi}_{n_4}\right)$ we obtain,

$$T\left(\hat{\varphi}_{n_1}\hat{\varphi}_{n_2}\hat{\varphi}_{n_3}\hat{\varphi}_{n_4}\right) \to : \hat{\varphi}_{n_1}\hat{\varphi}_{n_2}\hat{\varphi}_{n_3}\hat{\varphi}_{n_4} : + 6S_{n_1n_2} : \hat{\varphi}_{n_3}\hat{\varphi}_{n_4} : + 3S_{n_1n_2}S_{n_3n_4} \tag{7.6}$$

To arbitrary order,

$$T\left(\hat{\varphi}_{n_1}\hat{\varphi}_{n_2}\ldots\hat{\varphi}_{n_m}\right)\rightarrow\left(\begin{array}{l} :\hat{\varphi}_{n_1}\hat{\varphi}_{n_2}\ldots\hat{\varphi}_{n_m}: \\ +f_{m,1}:\hat{\varphi}_{n_1}\hat{\varphi}_{n_2}\ldots\hat{\varphi}_{n_{m-2}}:S_{n_{m-1},n_m} \\ +f_{m,2}:\hat{\varphi}_{n_1}\hat{\varphi}_{n_2}\ldots\hat{\varphi}_{n_{m-4}}:S_{n_{m-3},n_{m-2}}S_{n_{m-1},n_m} \\ \vdots \\ +\left\langle\begin{array}{l} f_{m,(m-1)/2}:\hat{\varphi}_{n_1}:S_{n_2,n_3}\ldots S_{n_{m-3},n_{m-2}}S_{n_{m-1},n_m}\text{ for } m=\text{odd} \\ \quad f_{m,m/2}S_{n_1,n_2}\ldots S_{n_{m-3},n_{m-2}}S_{n_{m-1},n_m}\text{ for } m=\text{even}\end{array}\right.\end{array}\right)\tag{7.7}$$

In the above $f_{m,r}$ is a combinatorial factor. For $f_{m,r}$ the index $m$ indicates the number of $\hat{\varphi}_n$ that the time order operator is operating on and the index $r$ indicates the number of pairings $S_{n,n'}$ . Two $\hat{\varphi}_n$ are used up for each $S_{n,n'}$ . The result is that,

$$f_{m,r}=\frac{m!}{r!(m-2r)!2^r}\tag{7.8}$$

Use (7.7) in (7.2) to obtain,

$$T\exp\left(iJ^n\hat{\varphi}_n\right)=1+\sum_{m=1}^{\infty}\frac{i^m}{m!}J^{n_1}J^{n_2}\ldots J^{n_m}\left(\begin{array}{l} :\hat{\varphi}_{n_1}\hat{\varphi}_{n_2}\ldots\hat{\varphi}_{n_m}: \\ +f_{m,1}:\hat{\varphi}_{n_1}\hat{\varphi}_{n_2}\ldots\hat{\varphi}_{n_{m-2}}:S_{n_{m-1},n_m} \\ +f_{m,2}:\hat{\varphi}_{n_1}\hat{\varphi}_{n_2}\ldots\hat{\varphi}_{n_{m-4}}:S_{n_{m-3},n_{m-2}}S_{n_{m-1},n_m} \\ \vdots \\ +\left\langle\begin{array}{l} f_{m,(m-1)/2}:\hat{\varphi}_{n_1}:S_{n_2,n_3}\ldots S_{n_{m-3},n_{m-2}}S_{n_{m-1},n_m}\text{ for } m=\text{odd} \\ \quad f_{m,m/2}S_{n_1,n_2}\ldots S_{n_{m-3},n_{m-2}}S_{n_{m-1},n_m}\text{ for } m=\text{even}\end{array}\right.\end{array}\right)\tag{7.9}$$

Next consider the right side of (7.1). In our condensed notation this is,

$$\hat{R}\equiv:\exp\left(iJ^n\hat{\varphi}_n\right):\exp\left(-\frac{i}{2}J^{n_1}J^{n_2}\Delta_{n_1n_2}\right)\tag{7.10}$$

Expand the exponentials to obtain,

$$\hat{R}=\sum_{k=0}^{\infty}\frac{i^k}{k!}:\left(J^n\hat{\varphi}_n\right)^k:\sum_{r=0}^{\infty}\frac{1}{r!}\left(-\frac{i}{2}\right)^r\left(J^{n_1}J^{n_2}\Delta_{n_1n_2}\right)^r\tag{7.11}$$

Use $i\Delta_F\left(x-x'\right)=T\left\langle 0\right|\hat{\varphi}_I\left(x\right)\hat{\varphi}_I\left(x'\right)\left|0\right\rangle$ so that $\Delta_{n_1n_2}=-iS_{n_1n_2}$ in the above to obtain,

$$\hat{R} = \left(1 + \sum_{k=1}^{\infty} \frac{i^k}{k!} :\left(J^n \hat{\varphi}_n\right)^k :\right)\left(1 + \sum_{r=1}^{\infty} \frac{i^{2r}}{r!2^r}\left(J^{n_1} J^{n_2} S_{n_1 n_2}\right)^r\right) \tag{7.12}$$

Rearrange terms to obtain,

$$\hat{R} = 1 + \sum_{r=1}^{\infty} C_{0,r} J^{n_1} J^{n_2} \ldots J^{n_{2r-1}} J^{n_{2r}} S_{n_1,n_2} \ldots S_{n_{2r-1},n_{2r}} + \sum_{k=1}^{\infty} C_{\mathrm{k},0} J^{n_1} J^{n_2} \ldots J^{n_k} : \hat{\varphi}_{n_1} \hat{\varphi}_{n_2} \ldots \hat{\varphi}_{n_k} : + \hat{W} \tag{7.13}$$

where,

$$C_{k,r} = \frac{i^{2r+k}}{k!r!2^r} \tag{7.14}$$

and,

$$\hat{W} = \sum_{k=1}^{\infty} \sum_{r=1}^{\infty} \begin{pmatrix} C_{k,r} J^{n_1} J^{n_2} \ldots J^{n_k} : \hat{\varphi}_{n_1} \hat{\varphi}_{n_2} \ldots \hat{\varphi}_{n_k} : \\ \times J^{n_{k+1}} J^{n_{k+2}} \ldots J^{n_{k+2r-1}} J^{n_{k+2r}} S_{n_1,n_2} \ldots S_{n_{k+2r-1},n_{k+2r}} \end{pmatrix} \tag{7.15}$$

Let $m = k + 2r$ which implies that $r = (m-k)/2$ . Change the summation to be over $m$ and $k$ instead of $k$ and $r$ to obtain,

$$\hat{W} = \sum_{m=1}^{\infty} J^{n_1} J^{n_2} \ldots J^{n_m} \sum_{k=1}^{m} \left(C_{k,(m-k)/2}\right)\left[: \hat{\varphi}_{n_1} \hat{\varphi}_{n_2} \ldots \hat{\varphi}_{n_k} : S_{n_{k+1},n_{k+2}} \ldots S_{n_{2m-1},n_{2m}}\right] \tag{7.16}$$

where,

$$C_{k,(m-k)/2} = \left\langle \begin{array}{l} \dfrac{i^k}{k!} \dfrac{i^{(m-k)}}{2^{[(m-k)/2]}\left[(m-k)/2\right]!} \quad \text{if } (m-k) \text{ is even or zero.} \\ \qquad\qquad 0 \qquad \text{if } k < 0 \text{ or } (m-k) \text{ is odd.} \end{array} \right. \tag{7.17}$$

Using the above we obtain,

$$\hat{R} = 1 + \sum_{m=1}^{\infty} J^{n_1} J^{n_2} \ldots J^{n_m} \left( \begin{array}{l} C_{m,0} : \hat{\varphi}_{n_1} \hat{\varphi}_{n_2} \ldots \hat{\varphi}_{n_m} : \\ + C_{m-2,1} : \hat{\varphi}_{n_1} \hat{\varphi}_{n_2} \ldots \hat{\varphi}_{n_{m-2}} : S_{n_{2m-1},n_{2m}} \\ + C_{m-4,2} : \hat{\varphi}_{n_1} \hat{\varphi}_{n_2} \ldots \hat{\varphi}_{n_{m-4}} : S_{n_{2m-1},n_{2m}} S_{n_{2m-1},n_{2m}} \\ \vdots \\ + \left\langle \begin{array}{l} m = odd, C_{1,[(m-1)/2]} : \hat{\varphi}_{n_1} : S_{n_2,n_3} \ldots S_{n_{2m-1},n_{2m}} S_{n_{2m-1},n_{2m}} \\ \quad m = even, C_{0,[m/2]} S_{n_1,n_2} \ldots S_{n_{2m-1},n_{2m}} S_{n_{2m-1},n_{2m}} \end{array} \right. \end{array} \right) \tag{7.18}$$

(7.1) is true if (7.18) equals (7.9). This will be true if,

$$\frac{i^m}{m!} f_{m,r} = C_{m-2r,r} \tag{7.19}$$

From (7.8) and (7.14) this is shown to be the case. Therefore (7.1) is true.